\documentstyle[epsfig,eqsecnum,aps]{revtex}
\begin{document}
\draft
\title{Counting pairs in $^{44}$Ti with various interactions}

\author{L. Zamick}
\address{ Department of Physics and Astronomy, Rutgers University, 
Piscataway, New Jersey USA 08855-8019}
\author{E. Moya de Guerra, P. Sarriguren, A. Escuderos}
\address{Instituto de Estructura de la Materia,
Consejo Superior de Investigaciones Cient\'{\i }ficas, \\
Serrano 123, E-28006 Madrid, Spain}
\author{ A.A. Raduta}
\address{Department of Theoretical Physics and Mathematics, 
Bucharest University, P.O.Box MG11,\\ and Institute of Physics and 
Nuclear Engineering, Bucharest, P.O.Box MG6, Romania}
\date{\today}
\maketitle

\begin{abstract}

We count the number of pairs in the single $j-$shell model of  
$^{44}$Ti for various interactions. For a state of total angular
momentum $I$, the wave function can be written as
$\Psi=\sum_{J_P\, J_N} D(J_P\, J_N) [(j^2)_{J_P}(j^2)_{J_N}]_I$,
where $D(J_P\, J_N)$ is the probability amplitude that the protons couple
to $J_P$ and the neutrons to $J_N$. For $I=0$ 
there are three states with ($I=0,\,T=0$) and one with ($I=0,\,T=2$). The
latter is the double analog of $^{44}$Ca. In that case ($T=2$), the magnitude 
of $D(JJ)$ is the same as that of a corresponding
two particle fractional parentage coefficient. In counting the number 
of pairs with an even angular momentum $J_A$ we find a new relationship
obtained by diagonalizing a unitary nine-j symbol.
We are also able to get results for the 'no-interaction' case for
$T=0$ states, for which it is found that there are less ($J=1,\,T=0$)
pairs than on the average. Relative to this 'no-interaction case'
we find for the most realistic interaction used that there
is an enhancement of pairs with angular momentum $J_A=0,2,1$ and 7, 
and a depletion for the others. Also considered are interactions in
which only the ($J=0,\,T=1$) pair state is at lower energy, only the 
($J=1,\,T=0$) pair state is lowered and where both are equally lowered.
\end{abstract}

%

\section{Introduction}
\label{sec:level1}
In this work we expand upon a note in a previous work on charge 
operator sum rules and proton-neutron $T=0$ and $T=1$ pairing 
interactions\cite{moya1}. The note in question has to do with
counting the number of pairs of particles with a given angular momentum
in $^{44}$Ti in a single $j-$shell approximation.

In the single $j-$shell model $^{44}$Ti consists of two valence protons
and two valence neutrons in the $f_{7/2}$ shell.
The allowed states for two identical particles have angular momenta
$J=0,2,4$ and 6 and isospin $T=1$. For a neutron-proton pair we can 
have these and also states of isospin $T=0$ with angular momenta
$J=1,3,5$ and 7. In other words, for even $J$ the isospin is one
and for odd $J$ the isospin is zero.

The wave function of a given state of total angular momentum $I$ can
be written as

\begin{equation}
\Psi=\sum_{J_P\, J_N} D^I(J_P,\, J_N) \left[ \left( j^2_\pi \right) ^{J_P}
\left( j^2_\nu \right) ^{J_N}\right] ^I\, .
\end{equation}
In the above $D(J_P,\, J_N)$ is the probability amplitude that the protons
couple to angular momentum $J_P$ and the neutrons to $J_N$. The normalization
condition is 

\begin{equation}
\sum_{J_P\, J_N} \left[ D^I(J_P,\, J_N)\right]  ^2=1,
\end{equation}
and the orthonormality condition is

\begin{equation}
\sum_{\alpha} D^\alpha(J_P,\, J_N) D^\alpha(J^\prime_P,\, J^\prime_N)
= \delta _{J_P,J^\prime_P} \delta_{J_N,J^\prime_N} .
\end{equation}

For states of angular momentum $I=0$, with which we will here be concerned,
$J_P$ must be equal to $J_N$ ($J=J_P=J_N$)

\begin{equation}
\Psi (I=0)=\sum_{J} D(JJ) \left[ \left( j^2_\pi \right) ^{J}
\left( j^2_\nu \right) ^{J}\right] ^0\, .
\end{equation}

In the single $j-$shell configuration of $^{44}$Ti there are three $I=0$
states of isospin $T=0$ and one of isospin $T=2$. The latter is the double
analog of a state in $^{44}$Ca, i.e., of a state of four neutrons in the
$f_{7/2}$ shell. For the unique ($I=0,\,T=2$) state in $^{44}$Ti, the 
magnitudes of the $D(JJ)$'s
are the same as those of two particle coefficients of fractional parentage

\begin{equation}
D(JJ)_{(I=0,\,T=2)}=\left( (j^2)_J (j^2)_J |\} j^40\right) .
\end{equation}
We thus have for ($I=0,\,T=2$)
\begin{equation}
D(00)=-0.5,\quad D(2,2)=0.3727,\quad D(4,4)=0.5,\quad D(6,6)=0.600
\end{equation}

For the ($I=0,\,T=0$) states however the $D$'s do depend on the interaction.
We show in Table 1 the values of the $D(JJ)$'s for the lowest energy
state for the following interactions

\begin {itemize}

\item A : ($J=0,\,T=1$) pairing (all two particle states are degenerate except 
($J=0,\,T=1$), which is lowered relative to the others.

\item B : ($J=1,\,T=0$) pairing. Only ($J=1,\,T=0$) is lowered.

\item C : Equal $J=0$ and $J=1$ pairing. Both ($J=0,\,T=1$) and ($J=1,\,T=0$) are
lowered by the same amount.

\item D :  MBZ interaction used by McCullen, Bayman and Zamick \cite{mbz}.

\item E : Spectrum of  $^{42}$Sc. This is the same as the MBZ calculation
except that the correct spectrum of  $^{42}$Sc is used (some of the $T=0$
states were not known in 1964). We equate the matrix elements
\begin{equation}
\left< \left( f_{7/2}\right) ^2_J \left|V \right| \left( f_{7/2}\right) ^2_J \right>
\end{equation}
with $E(J)$, the excitation energy of the lowest state of angular momentum
$J$ in $^{42}$Sc. The experimental values for $J=0$ to $J=7$ are (in MeV)
0.0, 0.6111, 1.5863, 1.4904, 2.8153, 1.5101, 3.2420, and 0.6163, respectively.
Note that the three lowest states have angular momenta $J=0,1,7$.

One can add to all those numbers a constant equal to the pairing energy
$E(^{42}Sc)+E(^{40}Ca)-E(^{41}Sc)-E(^{41}Ca)$. The value is $-3.182$ MeV.
However, adding this constant will not affect the spectrum or wave functions
of $^{44}$Ti.

Note that for the even-$J$ states of $^{42}$Sc the isospin is one, while
for the odd-$J$ states the isospin is zero.

The eigenvalues and eigenfunctions of interaction E are given in Table 2. 
\end{itemize}

\section{The number of pairs. A new relationship}
\label{secf:level2}

As previously noted\cite{moya1} for a system of nucleons with total
isospin $T$, we have the following result for the number of pairs:
\begin {itemize}
\item Total number of pair states is $n(n-1)/2$

\item Number with isospin $T_0=0$ is $n^2/8+n/4-T(T+1)/2$

\item Number with isospin $T_0=1$ is $3n^2/8-3n/4+T(T+1)/2$
\end{itemize}

Hence, for the $T=0$ state of $^{44}$Ti ($n=4$) we have three $T_0=0$ 
pairs and three $T_0=1$ pairs. For the $T=2$ state however we have six 
$T_0=1$ pairs. The important thing to note is that the number of pairs
does not depend on the two-body interaction, except for the fact that
it conserves isospin.

In $^{44}$Ti the number of pairs with total angular momentum $J_{12}$
($J_{12}=0,1,2,3,4,5,6,7)$ is given by

\begin{equation}
2\left| D(J_{12}J_{12}) \right| ^2 + \left| f(J_{12})\right| ^2\, ,
\end{equation}
where 

\begin{equation}
f\left( J_{12}\right) = 2 \sum_{J_P} U_{9j} (J_P,\, J_{12}) D(J_PJ_P)\, ,
\end{equation}
where we introduce the abbreviated symbol $U_{9j}$ to represent the unitary
$9j$ symbol,

\begin{eqnarray}
U_{9j}(J_PJ_{12})&=&\left< (j^2)J_P(j^2)J_P |(j^2)J_{12}(j^2)J_{12} \right> ^0 
\nonumber \\
&=&(2J_P+1)(2J_{12}+1)
\left\{ \matrix{j & j & J_P \cr j & j & J_P \cr J_{12} & J_{12} & 0} \right\}
\end{eqnarray}
A derivation of the results up to now in this section is given in Appendix A.
Since the last publication\cite{moya1} we have found a new relationship
which in some cases simplifies the expression. The new relationship
pertains to even $J_{12}$

\begin{eqnarray}
\sum_{J_P} U_{9j}(J_PJ_{12}) D(J_PJ_P) &=& D(J_{12}J_{12}) /2 \quad {\rm for}\quad 
T=0 \nonumber\\
&&= -D(J_{12}J_{12}) \quad {\rm for} \quad T=2
\end{eqnarray}
This remarkable relationship does not depend upon which isospin conserving
interaction is used. Using this result we find that the number of pairs for
even $J_{12}$ is equal to $3|D(J_{12},J_{12})|^2$. We do not have a 
corresponding simple expression for odd $J_{12}$.

We can prove the relationship by regarding the unitary 9j symbol
as a four by four matrix where $J_P$ and $J_{12}$ assume only even values
(0,2,4,6). The eigenvalues of this matrix are $-1$ (singly degenerate) and 
$0.5$ (triply degenerate). The eigenvalue $-1$ corresponds to the 
($J=0,\, T=2$) state of $^{44}$Ti and indeed the values of $D(JJ)$ are 
identical to those obtained with a charge independent Hamiltonian
$D(00)=-0.5,\, D(22)=0.3727,\, D(44)=0.5,\, D(66)=0.6009$. As previously
mentioned these are the two particle coefficients of fractional parentage.

The triple degeneracy with eigenvalue 0.5 corresponds to the three $T=0$ 
states being degenerate with this unitary 9j hamiltonian. This means that
any linear combination of the three $T=0$ states is an eigenvector. This
then proves the relationship.

\subsection{Results for the (I=0, T=2) state}

Since ($I=0,\,T=2$) state is unique, we will give the results for this case
first. Since the $^{44}$Ti $T=2$ state is the double analog of $^{44}$Ca,
a system of four identical particles, each pair must have even $J_{12}$.
The number of pairs is $6|<(j^2)J_{12}(j^2)J_{12}|\}j^40)|^2$, i.e., 
proportional to the square of the two particle coefficient fractional 
parentage. The number of pairs is:

1.5 for $J_{12}=0$; 0.8333 for $J_{12}=2$; 1.5 for $J_{12}=4$; and
2.16667  for $J_{12}=6$.

This is also the result for $^{44}$Ca. Hence, even though the $I=0$ ground state
of $^{44}$Ca has angular momentum zero and seniority zero, there are more
$J_{12}=6$ pairs in $^{44}$Ca than there are $J_{12}=0$ pairs.
This should not be surprising. As noted by Talmi \cite{talmi} for the
simpler case of a closed neutron shell i.e. $^{48}$Ca the number of pairs 
with angular momentum $J$ is
equal to $2J+1$. There is only one $J=0$ pair in  $^{48}$Ca.

\subsection{Number of pairs for all states}

We can count the number of pairs for all the four states, three of isospin
$T_{12}=0$ and one of isospin $T_{12}=1$.

Using the relation

\begin{equation}
\sum_\alpha D^\alpha (J_{12}J_{12}) D^\alpha (J^\prime _{12}J^\prime _{12})
=\delta_{J_{12}J^\prime_{12}} 
\end{equation}
we eliminate the $D$'s and find

\begin{equation}
({\rm Number\ of\ pairs}) /4 = \frac{1}{2} \delta_{J_{12}, even}+\frac{1}{2}
\left[ 1-U_{9j}(J_{12}J_{12})\right]
\end{equation}
The values for $T_{12}=0$ are 

0.9375 for $J_{12}=0$;  0.8542 for $J_{12}=2$; 0.9375 for  $J_{12}=4$;
and  1.0208 for $J_{12}=6$. The total sum is 3.75.

The values for $T_{12}=1$ are 

0.3244 for $J_{12}=1$; 0.6761 for $J_{12}=3$;  0.7494 for $J_{12}=5$;
and  0.5001 for $J_{12}=7$. The total sum is  2.25.

\subsection{Results for the T=0 ground state of $^{44}$Ti including
the no-interaction case}

In Table 3 we give results for the number of pairs for the four
interactions defined above. We also consider the 'no-interaction'
case. This is obtained by getting the total number of pairs for all
three $T=0$ states and dividing by three.

The results are given in Table 3. We start with the 'no-interaction'
result in the last column. Since there are six pairs and eight $J_{12}$'s
if there were an equal distribution, then we could assign 0.75 pairs
to each angular momentum. This serves us as a good basis for comparison.
We find that even in the 'no-interaction' case the results do depend
on $J_{12}$. The minimum number of pairs comes with the ($J_{12}=1,\,T_{12}=0$)
case, only 0.432. This is of interest because there has been a lot of
discussion in recent times about ($J_{12}=1,\,T_{12}=0$) pairing. We 
start out at least with a bias against it. The maximum number of pairs
in the 'no-interaction' case is for ($J_{12}=5,\,T_{12}=0$), a mode that
has largely been ignored.

However, of greater relevance is what happens to the ground state
wave function when the interaction is turned on. Therefore we compare
the no-interaction case with the {\it Spectrum of $^{42}$Sc} interaction case E. 
We see striking differences. Relative
to the no-interaction case there is an increase in the following number
of pairs: a) $J_{12}=0$ from 0.75 to 1.8617; b) $J_{12}=2$ from 0.861 to 0.9458;
c) $J_{12}=1$ from 0.432 to 0.6752 and d)$J_{12}=7$ from 0.667 to 1.8945.
Since the sum of all pairs in both cases is six, there must be a decrease
in the number of pairs with the other angular momenta and there is. For
example the number of  $J_{12}=6$ pairs decreases from 0.639 to 0.0457 and
the number of $J_{12}=5$ pairs from 1.00 to 0.1587. There are also large
decrease in the number of $J_{12}=4$ and $J_{12}=3$ pairs.
The results with MBZ are qualitatively similar to the correct spectrum
of $^{42}$Sc.

We next look at the schematic interactions in the first three columns.
For the ($J=0,\,T=1$) pairing interaction, there are a lot of $J_{12}=0$
pairs (2.25) but very few $J_{12}=1$ pairs (0.250). The number of
$J_{12}=7$ pairs is fairly large (1.250).

For the ($J=1,\,T=0$) pairing interaction there are a lot of $J_{12}=1$
pairs (1.297) and relatively few $J_{12}=0$ pairs (0.433). But still
there are substantial number of $J_{12}=7$ pairs (1.311). However, if we
examine the wave function for this case it is very different from that of
MBZ and this case represents a rather unrealistic ground state wave
function.

For the column equal  ($J=0,\,T=1$) and ($J=1,\,T=0$) pairing we get much 
better agreement in the wave function as compared with MBZ. The number of
$J_{12}=0$ pairs is 2.043 as compared with 1.736 from MBZ. For $J_{12}=1$
the values are 0.618 and 0.746, and for  $J_{12}=7$ are 1.654 and 1.948.
Amusingly when we lower the $J=0$ and $J=1$ matrix elements together
we get more $J_{12}=7$ pairs than we do for them separate (1.250 and 1.311).
There is one main deficiency in the $(J=0\,+\,J=1)$ case. The number of
($J=2,\,T=0$) pairs is only 0.497 as compared with 0.9458 for the {\it Spectrum
of $^{42}$Sc} case. The
enhancement is undoubtedly due to the quadrupole correlations in the
nucleus, an important ingredient which is sometimes forgotten when
all the emphasis is on ($J=0,\,T=1$) and ($J=1,\,T=0$) pairing. However 
if one restricts oneself to ($J=0,\,T=1$) and ($J=1,\,T=0$), then equal 
admixtures in the interaction yield much more realistic results than do
either one of them.

\section{Closing remarks}

In this work we have studied the effects of the nucleon-nucleon interaction
on the number of pairs of a given angular momentum in $^{44}$Ti. We have found
that the more attractive the nucleon-nucleon interaction is in a state with
angular momentum $J$, the more pairs of that given $J$ will be found in $^{44}$Ti.
As a basis of comparison we have defined the no-interaction case in which we
average over all three $T=0$ states (in the single $j-$shell approximation) in 
$^{44}$Ti, even if the number of $J$ pairs is not independent of $J$ and there
are for example less $J=1$ pairs than the average (0.432 vs. 6/8=0.75).
When the realistic interaction is turned on one gets, relative to this 
no-interaction case, an increase in the number of $J=0,1,2$ and 7 pairs and
a decrease in the others. This is in accord with the fact that in $^{42}$Sc
the states with angular momentum $J=0,1,2$ and 7 are lower than the others.

For the $T=2$ state of $^{44}$Ti, the double analog of $^{44}$Ca (a system
of particles of one kind), the number of $J=6$ pairs is the largest. This
suggests that the more deformed the state, (and the $T=0$ ground state of
$^{44}$Ti is certainly more deformed than the ground state of $^{44}$Ca), the
less the number of high angular momentum pairs with the exception of 
$J$(maximum)=7. Or to put it in another way, the more spherical the state
the higher the number of high angular momentum pairs.

The number of pairs obviously is of relevance to two nucleon transfer experiments
and we plan to address this more explicitly in the near future. For example,
the pickup of an $np$ pair in $^{44}$Ti to the $J=1$ state in $^{42}$Sc
will be enhanced, relative to the no-interaction case by a factor of
$(0.675/0.432)^2$. Of course $^{44}$Ti is not stable so this experiment cannot
be performed, so we will address other cases.

In the course of this work we also found a new relationship which simplifies the
expression for the number of even $J$ pairs
$$\sum_{J_B} D(J_BJ_B) <[(jj)J_B(jj)J_B]_0|[(jj)J_A(jj)J_A]_0> 
= D(J_AJ_A)/2$$ for T=0, 
and is equal to $-D(J_AJ_A)$ for $T=2$, where in the above
we have a unitary $9j$ symbol.

One way of looking at this is to say that we can also write the wave function
as 
$$
\sum_ {J,even} D(JJ) \left[ [p(1)n(2)] J [p(3)n(4)]\right]^{I=0} 
$$

\section{Appendix A. The number of pairs of a given angular momentum in the 
single $j-$shell in $^{44}$Ti }
\renewcommand{\theequation}{A.\arabic{equation}}
\setcounter{equation}{0}


The wave function of a given state in $^{44}$Ti is
\begin{equation}
\psi = \sum_{J_P J_N} D^I (J_P J_N) \left[ (j^2)^{J_P} (j^2)^{J_N} \right] ^I
\end{equation}
where $I$ is the total angular momentum.

In the single $j-$ shell there are 8 two-body interaction matrix elements
\begin{equation}
E(J) = \left< \left( f_{7/2}\right) ^2_J \left| V \right| 
\left( f_{7/2}\right) ^2_J \right>
\end{equation}
$J=0,1,...,7$. For even $J$ the isospin is $T=1$, for odd $J$ is $T=0$. The energy
of a $^{44}$Ti state can be written as $<\psi H \psi>$. This can be written
as a linear combination of the eight two-body matrix elements $E(J)$
\begin{equation}
E(^{44}Ti) = \sum_{J=0}^7 C_J E(J)
\end{equation}
We can identify $C(J)$ as the number of pairs in $^{44}$Ti with a given angular 
momentum $J$
\begin{equation}
<\psi H\psi> = \sum D(J'_PJ'_N)D(J_PJ_N) \left< \left[ J'_PJ'_N\right] ^I H \left[
J_PJ_N\right] ^I\right> 
\end{equation}

\begin{eqnarray}
&&\left< \left[ J'_PJ'_N\right] ^I H \left[ J_PJ_N\right] ^I\right> =
\left[ E\left( J_P\right) + E\left( J_N\right) \right] \delta_{J_PJ'_P}
\delta_{J_NJ'_N} \nonumber  \\
&&\qquad +4\sum_{J_AJ_B} \left< \left( j^2\right) J'_P  
\left( j^2\right) J'_N |  \left( j^2\right) J_A  \left( j^2\right) J_B
\right> ^I  \nonumber \\
&&\qquad \times \left< \left( j^2\right) J_P  
\left( j^2\right) J_N |  \left( j^2\right) J_A  \left( j^2\right) J_B
\right> ^I E(J_B)
\end{eqnarray}
In the above, the first two terms are the pp and nn interactions and the last 
one is the pn interaction. The factor of 4 is due to the fact that there are 4 np
pairs. The unitary 9j symbol recombines a proton and a neutron.
Note that $J_P$ and $J_N$ are even but $J_A$ and $J_B$ can be even or odd.

By identifying the coefficient of $E(J_B)$ as the number of pairs with
angular momentum $J_B$ we get the expression for $I=0$ (for which $J_P=J_N$)

\begin{eqnarray}
&&{\rm Number\; of}\; J_B\; {\rm pairs} = 2D^\alpha (J_PJ_P)_{(J_P\; even)}
\nonumber \\
&&\qquad +4\sum_{J_PJ_N}  D(J_PJ_N) \left< \left( j^2\right) J_P  
\left( j^2\right) J_N |  \left( j^2\right) J_B \left( j^2\right) J_B
\right> ^0 \nonumber \\
&&\qquad \times \sum_{J'_PJ'_N} D(J'_PJ'_N)
 \left< \left( j^2\right) J'_P  
\left( j^2\right) J'_N |  \left( j^2\right) J_B  \left( j^2\right) J_B
\right> ^0 
\end{eqnarray}
The first term contributes only for $J_B$ even.

We can rewrite this as 
\begin{equation}
{\rm Number\; of}\; J_B\; {\rm pairs} = 2D^\alpha (J_PJ_P)_{(J_P\; even)}\quad
+\left| f(J_B)\right| ^2
\end{equation}
with
\begin{equation}
f(J_B)= 2  \sum_{J_PJ_N} D(J_PJ_N) \left< \left( j^2\right) J_P  
\left( j^2\right) J_N |  \left( j^2\right) J_B  \left( j^2\right) J_B
\right> ^0
\end{equation}

\begin{eqnarray}
&&\left< \left( j_1j_2\right) J_P  
\left( j_3j_4\right) J_N |  \left( j_1j_3\right) J_A  \left( j_2j_4\right) J_B
\right> ^I \nonumber \\
&&=\sqrt{(2J_P+1)(2J_N+1)(2J_A+1)(2J_B+1)}
\left\{ \matrix{j_1 & j_2 & J_P \cr j_3 & j_4 & J_N \cr J_A & J_B & I} \right\}
\end{eqnarray}

We can get the number of pairs with a given isospin $T$ by using a simple
interaction $a+b t(1)\cdot t(2)$, where $a$ and $b$ are constants. The value
of this interaction for two particles is
\begin{equation}
a-3b/4 \; ({\rm for} \; T_0=0) \quad {\rm and}\quad
a+b/4 \; ({\rm for} \; T_0=1)
\end{equation}
For $n$ nucleons we have
\begin{eqnarray}
\sum_{i<j} \left( a+b t(i)\cdot t(j)\right) &=&
\frac{a}{2}n(n-1) +\frac{b}{2}\sum_{i,j}t(i)\cdot t(j)
-\frac{1}{2}\sum_{i} t(i)^2 \nonumber \\
&=&\frac{a}{2}n(n-1)+\frac{b}{2}T(T+1) -\frac{3}{8}nb
\end{eqnarray}
We can write this as 
\begin{equation}
C_0 (a-3b/4)+C_1 (a+b/4)
\end{equation}
and identify $C_T$ as the number of pairs with isospin $T$. We then get the result
of section 2.
\begin{eqnarray}
C_0 &=& \frac{n^2}{8}+\frac{n}{4}-\frac{T(T+1)}{2} \\
C_1 &=& \frac{3n^2}{8}-\frac{3n}{4}+\frac{T(T+1)}{2}
\end{eqnarray}

\begin{center}
{\Large \bf Acknowledgments} 
\end{center}

This work was partially supported by MCyT (Spain) under contract number
BFM2002-03562, by  a U.S. Dept. of Energy Grant No. DE-FG02-95ER-40940 and by 
a NATO Linkage Grant PST 978158.

\vfill\eject

\vfill \eject

\begin{table}[h]

{\bf Table 1.} Wave functions of $^{44}$Ti for various interactions:
A ($J=0,\,T=1$) pairing;  B ($J=1,\,T=0$) pairing;  
C equal $J=0,J=1$ pairing; D (MBZ);  E spectrum of $^{42}$Sc.
\vskip .5cm

\begin{center}
\begin{tabular}{ccccccc}
\hline
&\multicolumn{5}{c} {Ground state $T=0$}& $T=2$ \\
\cline{2-6}
 $D(JJ)$ & A & B & C & D & E & any interaction  \\
\hline
$J=0$  & 0.866 &  0.380 & 0.826 & 0.7608 & 0.7878 & -0.5   \\
$J=2$  & 0.213 &  0.688 & 0.405 & 0.6090 & 0.5617 & 0.3727 \\            
$J=4$  & 0.289 &  0.416 & 0.373 & 0.2093 & 0.2208 & 0.5    \\  
$J=6$  & 0.347 & -0.457 & 0.126 & 0.0812 & 0.1234 & 0.6009 \\      
\hline
\end{tabular}
\end{center}
\end{table}

\vskip 1cm

\begin{table}[h]

{\bf Table 2.} 
Excitation energies [MeV] and eigenvectors of the {\it Spectrum of $^{42}$Sc} 
interaction.

\vskip .5cm

\begin{center}
\begin{tabular}{ccccc}
\hline
excitation energies & \multicolumn{4}{c} {eigenvectors} \\
\cline{2-5}
& D(00) & D(22) & D(44) & D(66) \\
0 & 0.78776 & 0.56165 & 0.22082 & 0.12341 \\
5.5861 &-0.35240 & 0.73700 &-0.37028 &-0.44219 \\
8.2840 &-0.50000 & 0.37268 & 0.50000 & 0.60093 \\
8.7875 &-0.07248 &-0.04988 & 0.75109 &-0.65432 \\
\hline
\end{tabular}
\end{center}
\end{table}

\vskip 1cm

\begin{table}[h]

{\bf Table 3.} 
Number of pairs for various interactions:
A ($J=0,\,T=1$) pairing;  B ($J=1,\,T=0$) pairing;  
C equal $J=0,J=1$ pairing; D (MBZ); E spectrum of $^{42}$Sc;
F no interaction.

\vskip .5cm

\begin{center}
\begin{tabular}{ccccccc}
\hline
& A & B & C & D & E & F \\

$J_{12}=0$ & 2.250 & 0.433 & 2.045 & 1.736 & 1.862 & 0.750 \\ 
$J_{12}=2$ & 0.139 & 1.420 & 0.492 & 1.126 & 0.946 & 0.861 \\
$J_{12}=4$ & 0.250 & 0.320 & 0.416 & 0.114 & 0.146 & 0.750 \\
$J_{12}=6$ & 0.361 & 0.626 & 0.048 & 0.020 & 0.046 & 0.639 \\ \\
$J_{12}=1$ & 0.250 & 1.297 & 0.618 & 0.746 & 0.675 & 0.432 \\
$J_{12}=3$ & 0.583 & 0.388 & 0.165 & 0.216 & 0.271 & 0.902 \\
$J_{12}=5$ & 0.916 & 0.003 & 0.564 & 0.091 & 0.159 & 1.000 \\
$J_{12}=7$ & 1.250 & 1.311 & 1.654 & 1.948 & 1.895 & 0.667 \\
\hline
\end{tabular}
\end{center}
\end{table}

\vskip 1cm

\end{document}